\documentclass[12pt]{article}

\usepackage{graphicx}
\usepackage{amsmath}

\def\ltwid{\mathrel{\raise.3ex\hbox{$<$\kern-.75em\lower1ex\hbox{$\sim$}}}}
\def\gtwid{\mathrel{\raise.3ex\hbox{$>$\kern-.75em\lower1ex\hbox{$\sim$}}}}

\def\square{\kern1pt\vbox{\hrule height 1.2pt\hbox{\vrule width 1.2pt\hskip 3pt
   \vbox{\vskip 6pt}\hskip 3pt\vrule width 0.6pt}\hrule height 0.6pt}\kern1pt}
\def\overleftrightarrow#1{\vbox{\ialign{##\crcr
     $\leftrightarrow$\crcr\noalign{\kern-1pt\nointerlineskip}
     $\hfil\displaystyle{#1}\hfil$\crcr}}}

\begin{document}

\begin{titlepage}

\begin{flushright}
ITP-UU-12/31, SPIN-12/29, UFIFT-QG-12-06
\end{flushright}

\vskip 1cm

\begin{center}
{\bf Covariant Vacuum Polarizations on de Sitter Background}
\end{center}

\vskip .5cm

\begin{center}
Katie E. Leonard$^{1*}$, T. Prokopec$^{2\dagger}$ and R. P.
Woodard$^{1,\ddagger}$
\end{center}

\vskip .5cm

\begin{center}
\it{$^{1}$ Department of Physics, University of Florida \\
Gainesville, FL 32611, UNITED STATES}
\end{center}

\begin{center}
\it{$^{2}$ Institute of Theoretical Physics (ITP) \& Spinoza Institute, \\
Utrecht University, Postbus 80195, 3508 TD Utrecht, \\ THE NETHERLANDS}
\end{center}

\vskip .5cm

\begin{center}
ABSTRACT
\end{center}

We derive covariant expressions for the one loop vacuum polarization
induced by a charged scalar on de Sitter background. Two forms are
employed: one in which two covariant derivatives act on de Sitter
invariant basis tensors multiplied by scalar structure functions,
and the other in which four covariant derivatives act on a single
basis tensor times a structure function. The second representation
permits the correction to dynamical photons to be expressed as a
surface integral, which raises the important question of what sorts
of effects can be absorbed into corrections of the initial state.
Results are obtained for charged, minimally coupled scalars which
are either massless or else light. The former show de Sitter
breaking whereas the latter are de Sitter invariant. However, the de
Sitter invariant formulation does not seem useful, even when it is
possible. Our work has important implications for representations of
the graviton self-energy on de Sitter.

\begin{flushleft}
\end{flushleft}

\vskip .5cm

\begin{flushleft}
$^*$ e-mail: katie@phys.ufl.edu \\
$^{\dagger}$ e-mail: T.Prokopec@uu.nl \\
$^{\ddagger}$ e-mail: woodard@phys.ufl.edu
\end{flushleft}

\end{titlepage}

\section{Introduction}\label{intro}

The motivation for this paper is finding the most effective
representation for the vacuum polarization $i[\mbox{}^{\mu}
\Pi^{\nu}](x;x')$ from scalar quantum electrodynamics (SQED) on de
Sitter background \cite{PTW,PP}. This quantity is notable because it
provides a mechanism by which the photon can gain a mass during
inflation \cite{PW}. It has been suggested that a residual effect
from this mass generation might be the origin of primordial magnetic
fields \cite{DDPT}. However, we will here focus only on how to
represent previous results for the vacuum polarization \cite{PTW,PP}
as bitensor functions of space and time.

The context is the long controversy over which is the more powerful
organizing principle for quantum field theory on de Sitter
background: the conformal flatness of de Sitter \cite{TW1,RPW,MTW4},
or the manifold's isometry group \cite{Higuchi,HMM}. Classical
electromagnetism is conformally invariant in $D=4$ dimensions, which
reduces the system to its well understood flat space limit. This
motivated the decision to exploit conformal flatness in representing
the one loop vacuum polarization from a massless, minimally coupled
scalar \cite{PTW}, and from a light scalar \cite{PP}. However, there
has never been a de Sitter invariant form with which to compare. And
it is known that both the one loop effective potential and the two
loop expectation values of scalar and field strength bilinears
\cite{PTsW} are simplified by employing a de Sitter invariant gauge
for the photon propagator \cite{TW2}.

Our goal is to re-express the old results \cite{PTW,PP} for the
vacuum polarization in a form consisting of covariant derivatives
acting on a scalar structure function, much like the familiar flat
space representation,
\begin{equation}
\Bigl[ \mbox{}^{\mu} \Pi^{\nu}\Bigr]_{\rm flat}\!\!\!\!\!\!\!(x;x')
= \Bigl[\eta^{\mu\nu} \partial^2 \!-\! \partial^{\mu}
\partial^{\nu}\Bigr] \Pi\Bigl( (x \!-\! x')^2 \Bigr) \; .
\label{flatvac}
\end{equation}
The form we employ is guaranteed to be manifestly de Sitter
invariant if the vacuum polarization is physically invariant. The
massless, minimally coupled scalar shows a physical breaking of de
Sitter invariance \cite{AF}, and our covariant representation
reveals that this breaking afflicts the one loop vacuum
polarization. The massive scalar has a de Sitter invariant
propagator so we obtain a fully invariant result for its structure
function. In neither case is the result particularly illuminating.
Indeed, the de Sitter invariant formalism seems to obscure the
essential physics, but this could not have been known beforehand.

\begin{figure}[!ht]
\centering
\includegraphics[scale = 0.45]{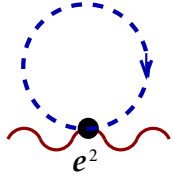}
\caption{\label{fig1} One loop contribution to the vacuum
polarization from the 4-point (seagull) interaction.}
\end{figure}

\begin{figure}[!ht]
\centering
\includegraphics[scale = 0.35]{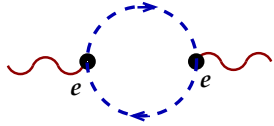}
\caption{\label{fig2} One loop contribution to the vacuum
polarization from the 3-point interaction.}
\end{figure}

\begin{figure}[!ht]
\centering
\includegraphics[scale = 0.45]{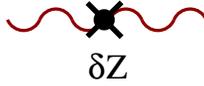}
\caption{\label{fig3} Photon field strength renormalization
counterterm.}
\end{figure}

The vacuum polarization we wish to calculate comes from the three
diagrams shown in Figs.~\ref{fig1}-\ref{fig3}.
We represent the de Sitter background geometry in open conformal
coordinates, which is conducive to regarding it as a limiting form
of primordial inflation. The invariant element is
\begin{equation}
ds^2 = a^2 \left[-d\eta^2 + d\vec{x} \cdot d\vec{x} \right] \; ,
\end{equation}
where $a(\eta)=-\frac{1}{H\eta}=e^{Ht}$ is the scale factor and $H$
is the Hubble parameter. Gravity is treated as a non-dynamical
background, with the vector potential $A_\mu(x)$ and the complex
scalar $\phi(x)$ being the dynamical variables. The Lagrangian
describing massless SQED is
\begin{equation}
\mathcal{L}_{m=0}=-\frac{1}{4}F_{\mu\nu}F_{\rho\sigma}
g^{\mu\rho} g^{\nu\sigma} \sqrt{-g} - (\partial_\mu -ie A_\mu)
\phi^*(\partial_\nu + ie A_\nu) \phi g^{\mu\nu}\sqrt{-g} \; ,
\end{equation}
where $F_{\mu\nu}\equiv \partial_\mu A_\nu-\partial_\nu A_\mu$
is the field strength. The Lagrangian for massive SQED is
\begin{equation}
\mathcal{L}_{m\neq 0}=\mathcal{L}_{m=0} - M^2 \phi^* \phi \sqrt{-g}
\; ,
\end{equation}
where we are particularly interested in the case $M \ll H$.

In representing functions which depend upon two points, $x^{\mu}$
and ${x'}^{\mu}$, we will make extensive use of the de Sitter length
function
\begin{equation}
y(x;x') \equiv a(\eta) a(\eta') H^2 \left[|| \vec{x} - \vec{x}'||^2
- (| \eta - \eta'| - i \delta)^2 \right] \; .
\end{equation}
We also have need of two de Sitter breaking combinations of the
scale factor $a$ at $x^{\mu}$ and $a'$ at ${x'}^{\mu}$
\begin{equation}
u \equiv \ln(aa') \; , \quad {\rm and} \quad v \equiv
\ln\Bigl(\frac{a}{a'}\Bigr) \; .
\end{equation}
Derivatives of $y$ and $u$ furnish a convenient basis for
representing bi-vector functions of $x^{\mu}$ and ${x'}^{\mu}$ such
as the vacuum polarization
\begin{equation}
\partial_{\mu} y \quad , \quad \partial'_{\nu} y \quad , \quad
\partial_{\mu} \partial'_{\nu} y \quad , \quad \partial_{\mu} u
\quad , \quad \partial'_{\nu} u \; . \label{basis}
\end{equation}
We do not need derivatives of $v$ because they are related to those
of $u$
\begin{equation}
\partial_{\mu} v = \partial_{\mu} u \qquad , \qquad \partial'_{\nu}
v = - \partial'_{\nu} u \; .
\end{equation}
It turns out that either taking covariant derivatives of any of the
five basis tensors (\ref{basis}), or contracting any two of them
into one another, produces metrics and more basis tensors
\cite{KW,MTW2}.

Section~\ref{MMC} develops a representation, based on two covariant
derivatives, for the vacuum polarization from a massless, minimally
coupled scalar \cite{PTW}. In this case there is de Sitter breaking,
which requires two structure functions. When there is no de Sitter
breaking the vacuum polarization can be expressed in terms of just a
single structure function using a representation which involves four
covariant derivatives. Section~\ref{mneq0} derives this
representation for the case of a massive scalar \cite{PP}. In
section~\ref{dsinv} we use the result of section~\ref{mneq0} to
study the effective field equations. An especially counter-intuitive
and confusing feature of the de Sitter invariant representation is
that local corrections to the effective field equations manifest as
surface terms from the initial time. Section~\ref{discuss} comprises
our discussion.

\section{A Massless, Minimally Coupled Scalar}\label{MMC}

The purpose of this section is to express the one loop vacuum
polarization from a massless, minimally coupled scalar using a
neutral representation that would be manifestly de Sitter invariant
if the physics was. In the first sub-section we review the primitive
expression for the three diagrams of Figs.~\ref{fig1}-\ref{fig3}.
The next sub-section derives the most general transverse form the
vacuum polarization can take consistent with the symmetries of
cosmology. In the final sub-section we give a simple solution for
the structure functions which reproduces the primitive result.

\subsection{The Primitive Result}

If we call the scalar propagator $i\Delta(x;x')$ then the three
diagrams in Figs.~\ref{fig1}-\ref{fig3} make the following
contribution to the vacuum polarization
\begin{eqnarray}
\label{orig_vacpol} \lefteqn{i\Bigl[\mbox{}^\mu \Pi^\nu \Bigr](x;x')
= -2i e^2 \sqrt{-g} \, g^{\mu\nu} i\Delta(x;x) \delta^D(x \!-\! x')
} \nonumber \\
& & \hspace{-.5cm} + 2 e^2 \sqrt{-g} \, g^{\mu\rho} \sqrt{-g'} \,
g'^{\nu\sigma} \Bigl[\partial_\rho i\Delta(x;x')
\partial'_\sigma i\Delta(x;x') \!-\! i \Delta(x;x')
\partial_\rho \partial'_\sigma i\Delta(x;x') \Bigr] \nonumber \\
& & \hspace{3.5cm} +i\delta Z \partial_{\rho} \Bigl[ \sqrt{-g}
\Bigl( g^{\mu\nu} g^{\rho\sigma} \!-\! g^{\mu\sigma} g^{\nu\rho}
\Bigr) \partial_{\sigma} \delta^D(x-x') \Bigr] \; . \qquad
\end{eqnarray}
Equation (\ref{orig_vacpol}) is valid for any scalar. For the
special case of the massless, minimally coupled scalar the
propagator obeys
\begin{equation}
\square i\Delta_A(x;x') = \frac{i \delta^D(x-x')}{\sqrt{-g}} \; ,
\label{Apropeqn}
\end{equation}
where $\square \equiv (-g)^{-\frac12} \partial_{\mu} (\sqrt{-g} \,
g^{\mu\nu} \partial_{\nu})$ is the covariant scalar d'Alembertian.
It has long been known that equation (\ref{Apropeqn}) has no de
Sitter invariant solution \cite{AF}, so the scalar propagator must
break some of the de Sitter symmetries. When using de Sitter as a
paradigm for primordial inflation one obviously wishes to preserve
the symmetries of cosmology: homogeneity, isotropy and spatial
flatness. In that case the unique solution takes the form \cite{OW}
\begin{equation}
i \Delta_A(x;x') = A\Bigl( y(x;x') \Bigr) + k \ln(a a') = A(y) + k u
\; ,
\end{equation}
where the constant $k$ is
\begin{equation}
k = \frac{H^{D-2}}{(4\pi)^{\frac{D}2}} \frac{\Gamma(D
\!-\!1)}{\Gamma(\frac{D}{2})} \; .
\end{equation}
The de Sitter invariant part of the scalar propagator is
\begin{eqnarray}
\lefteqn{A(y) = \frac{H^{D-2}}{(4\pi)^{\frac{D}2}} \Biggl\{
\frac{\Gamma(\frac{D}2)}{\frac{D}2 \!-\! 1}
\Bigl(\frac{4}{y}\Bigr)^{\frac{D}2 -1} + \frac{\Gamma(\frac{D}2
\!+\! 1)}{\frac{D}2 \!-\! 2} \Bigl(\frac{4}{y}\Bigr)^{\frac{D}2 -2}
 } \nonumber \\
& & \hspace{.5cm} -\displaystyle\sum\limits_{n=1}^\infty \Biggl[
\frac{\Gamma(n \!+\! \frac{D}2 \!+\! 1)}{(n \!-\! \frac{D}2 \!+\!
2)(n \!+\! 1)!} \Bigl(\frac{y}{4}\Bigr)^{n -\frac{D}2 +2} -
\frac{\Gamma(n \!+\! D \!-\! 1)}{n \Gamma(n \!+\! \frac{D}2)}
\Bigl(\frac{y}{4}\Bigr)^{n} \Biggr] \Biggr\}
+ A_1
 , \qquad \label{Aexp}
\end{eqnarray}
where $A_1$ is a D-dependent constant which diverges for $D = 4$
\begin{equation}
A_1 = \frac{H^{D-2}}{(4\pi)^{D/2}}\frac{\Gamma(D \!-\! 1)}{\Gamma(\frac{D}2)}
\Biggl\{-\psi\Bigl(1 \!-\! \frac{D}2\Bigr) \!+\! \psi\Bigl(\frac{D
\!-\! 1}{2} \Bigr) \!+\! \psi(D \!-\! 1) \!+\! \psi(1) \Biggr\} \; .
\end{equation}
As a consequence of the propagator equation (\ref{Apropeqn}) the
function $A(y)$ obeys
\begin{equation}
(4 y \!-\! y^2) A''(y) + D (2 \!-\! y) A'(y) = (D \!-\! 1) k \; .
\label{Aeqn}
\end{equation}

Derivatives of the propagator can be expressed using the tensor
basis (\ref{basis})
\begin{eqnarray}
\partial_{\mu} i\Delta_A(x;x') & = & A'(y) \times \partial_{\mu} y + k
\times \partial_{\mu} u \; , \\
\partial'_{\nu} i\Delta_A(x;x') & = & A'(y) \times \partial'_{\nu} y + k
\times \partial'_{\nu} u \; , \\
\partial_{\mu} \partial'_{\nu} i\Delta_A(x;x') & = & A'(y) \times
\partial_{\mu} \partial'_{\nu} y + A''(y) \times \partial_{\mu} y \,
\partial'_{\nu} y \nonumber \\
& & \hspace{4cm} + \frac{i \delta^D(x \!-\! x')}{H^2 \sqrt{-g} }
\times \partial_{\mu} u \, \partial'_{\nu} u \; . \qquad
\end{eqnarray}
By applying these identities we can write (\ref{orig_vacpol}) as
\begin{eqnarray}
\lefteqn{i\Bigl[\mbox{}^\mu \Pi^\nu \Bigr](x;x') = -2i e^2 \sqrt{-g}
\, g^{\mu\rho} g^{\nu\sigma} \Bigl[ g_{\rho\sigma} \!+\! H^{-2}
\partial_{\rho} u \partial_{\sigma} u\Bigr] (A_1 \!+\! k u)
\delta^D(x \!-\! x') } \nonumber \\
& & \hspace{-.5cm} + 2 e^2 \sqrt{-g} \, g^{\mu\rho} \sqrt{-g'} \,
g'^{\nu\sigma} \Biggl\{ \partial_\rho y\partial'_\sigma y \!\times\!
\Bigl(A'^2 \!-\! A A'' \!-\! k u A'' \Bigr) \nonumber \\
& & \hspace{0cm} - \partial_\rho \partial'_\sigma y \!\times\!
\Bigl(A A' \!+\! k u A' \Bigr) \!+\! \partial_\rho u
\partial'_\sigma y \times k A' \!+\! \partial_\rho y
\partial'_\sigma u \!\times \! k A' \!+\! \partial_\rho u
\partial'_\sigma u \!\times\! k^2 \Biggr\} \nonumber \\
& & \hspace{3.5cm} +i\delta Z \partial_{\rho} \Bigl[ \sqrt{-g}
\Bigl( g^{\mu\nu} g^{\rho\sigma} \!-\! g^{\mu\sigma} g^{\nu\rho}
\Bigr) \partial_{\sigma} \delta^D(x-x') \Bigr] \; . \qquad
\label{old_vacpol}
\end{eqnarray}
The presence of de Sitter breaking in (\ref{old_vacpol}) is evident
from the dependence upon $u$ and its derivatives.

\subsection{General Representations}

We wish to express (\ref{old_vacpol}) using a manifestly transverse
representation analogous to the form (\ref{flatvac}) used for flat
space. An important property of this representation is that it
should be ``neutral''. That is, it should be in terms of covariant
derivatives so that the structure functions will be de Sitter
invariant if the physical result happens to possess that symmetry.
We are therefore led to the ansatz
\begin{equation}
\label{ansatz} i \Bigl[\mbox{}^\mu \Pi^\nu \Bigr](x;x') =
\sqrt{-g(x)} \, \sqrt{-g(x')} \, D_\rho D'_\sigma
\Bigl[\mbox{}^{\mu\rho} T^{\nu\sigma} \Bigr](x;x') \; ,
\end{equation}
where the bi-tensor $[\mbox{}^{\mu\rho}T^{\nu\sigma}](x;x')$
must be antisymmetric under $\mu \leftrightarrow \rho$ and $\nu
\leftrightarrow \sigma$. It must also obey the ``reflection
identity''
\begin{equation}
\Bigl[\mbox{}^{\mu\rho} T^{\nu\sigma}\Bigr](x;x') =
\Bigl[\mbox{}^{\nu\sigma} T^{\mu\rho}\Bigr](x';x) \; .
\end{equation}
The most general tensor satisfying these criteria, consistent with
homogeneity and isotropy, is
\begin{eqnarray}
\label{generic_tf} \lefteqn{\Bigl[\mbox{}_{\mu\rho} T_{\nu\sigma} \Bigr]
= \partial_\mu \partial'_{[\nu} y \, \partial'_{\sigma]} \partial_\rho y
\!\times\! f_1 + \partial_{[\mu} y \, \partial_{\rho]} \partial'_{[\nu} y \,
\partial'_{\sigma]} y \!\times\! f_2 + \partial_{[\mu} y \, \partial_{\rho]}
\partial'_{[\nu} y \, \partial'_{\sigma]} u \!\times\! f_3} \nonumber \\
& & \hspace{-.5cm} + \partial_{[\mu}u \, \partial_{\rho]}
\partial'_{[\nu} y \partial'_{\sigma]} y \!\times\! \tilde{f}_3 +
\partial_{[\mu}u \, \partial_{\rho]} \partial'_{[\nu}y \,
\partial'_{\sigma]} u \!\times\! f_4 + \partial_{[\mu}y \, \partial_{\rho]}
u \, \partial'_{[\nu}y \partial'_{\sigma]}u \!\times\! f_5 \; . \qquad
\end{eqnarray}
Here the scalar structure functions $f_i(y,u,v)$ depend on $y$, $u$
and $v$, and we define $\tilde{f}_3(y,u,v) \equiv f_3(y,u,-v)$. Of
course manifest de Sitter invariance precludes $f_{3},f_{4}$ and $f_{5}$, 
and we will in the end employ only $f_1$ and $f_2$.

There is obviously some redundancy in the five structure functions
of (\ref{generic_tf}) because acting the derivatives in
(\ref{ansatz}) results in only four algebraically independent
tensors,
\begin{eqnarray}
\lefteqn{ D^{\rho} {D'}^{\sigma} \Bigl[ \mbox{}_{\mu\rho}
T_{\nu\sigma} \Bigr](x;x') = \partial_{\mu} \partial'_{\nu} y
\!\times\! F_1 \!+\! \partial_{\mu} y \, \partial'_{\nu} y
\!\times\! F_2} \nonumber \\
& & \hspace{4cm} + \partial_{\mu} u \, \partial'_{\nu} y \!\times\!
F_3 \!+\! \partial_{\mu} y \, \partial'_{\nu} u \!\times\!
\widetilde{F}_3 \!+\! \partial_{\mu} u \, \partial'_{\nu} u
\!\times\! F_4 \; . \qquad \label{Dfs}
\end{eqnarray}
Conservation implies two differential relations \cite{MTW2},
\begin{eqnarray}
\lefteqn{ \Bigl[ -(2\!-y) \partial_y \!+\! D \!+\! \partial_u \!+\!
\partial_v \!+\! 2 e^{-v} \partial_y \Bigr] F_3 = \Bigl[ (2 \!-\! y)
\partial_y \!-\! D \!-\! \partial_u \!-\! \partial_v \Bigr] F_1 }
\nonumber \\
& & \hspace{1.8cm} + \Bigl[ (4y \!-\! y^2) \partial_y \!+\! (D \!+\!
1) (2 \!-\! y) \!+\! (2 \!-\! y \!-\! 2 e^{-v} )
(\partial_u \!+\! \partial_v) \Bigr] F_2 \; , \qquad \label{cons1} \\
\lefteqn{ \Bigl[ -(2 \!-\! y) \partial_y \!+\! (D \!-\! 1) \!+\!
\partial_u \!+\! \partial_v \!+\! 2 e^{-v} \partial_y \Bigr] F_4 =
-2 e^{-v} (\partial_u \!+\! \partial_v) F_1 \!-\! 2 e^{-v} F_3 }
\nonumber \\
& & \hspace{2.7cm} + \Bigl[ (4 y \!-\! y^2) \partial_y \!+\! D (2
\!-\! y) \!+\! (2 \!-\! y \!-\! 2 e^{-v} ) (\partial_u \!+\!
\partial_v )\Bigr] \widetilde{F}_3 \; . \qquad \label{cons2}
\end{eqnarray}
This suggests that one should be able to write the scalar
coefficient functions $F_i(y,u,v)$ of expression (\ref{Dfs}) in
terms of just two master structure functions, which we might call
$\Phi(y,u,v)$ and $\Psi(y,u,v)$. After long reflection, one sees
that the following substitutions for the $F_i(y,u,v)$ reduce the
conservation relations (\ref{cons1}-\ref{cons2}) to tautologies,
\begin{eqnarray}
F_1 \!\!\! & = & \!\!\! \Bigl[ -(4 y \!-\! y^2)
\partial_y \!-\! (D\!-\!1) (2 \!-\! y) \!-\! 2 (2 \!-\! y)
\partial_u \!+\! 4 \cosh(v) \partial_u \!-\! 4 \sinh(v) \partial_v
\Bigr] \Phi \nonumber \\
& & \hspace{6.5cm} + \Bigl[4 \cosh(v) \!-\! (2 \!-\! y) \Bigr]
\Psi \!-\! \Xi \; , \qquad \label{F1} \\
F_2 \!\!\! & = & \!\!\! \Bigl[ (2 \!-\! y)
\partial_y \!-\! D \!+\! 1 \!-\! 2 \partial_u \Big] \Phi \!-\! \Psi
\; , \qquad \label{F2} \\
F_3 \!\!\! & = & \!\!\! -2 e^{v} (\partial_u \!-\!
\partial_v) \Phi \!-\! 2 e^{v} \Psi \!+\! \Xi \; , \label{F3} \\
\widetilde{F}_3 \!\!\! & = & \!\!\! -2 e^{-v} (\partial_u \!+\!
\partial_v) \Phi \!-\! 2 e^{-v} \Psi \!+\! \Xi \;
, \label{F3tilde} \\
F_4 \!\!\! & = & \!\!\! -4 \Psi \!+\! (2 \!-\! y) \Xi \; .
\label{F4}
\end{eqnarray}
Here the auxiliary function $\Xi(y,u,v)$ is,
\begin{equation}
\Xi(y,u,v) \equiv \int \!\! dy \, (\partial_u^2 \!-\! \partial_v^2)
\Phi(y,u,v) \!+\! (2 \!-\! y) \Psi(y,u,v) \!-\! (D \!-\! 2) \!\!
\int \!\! dy \, \Psi(y,u,v) \; . \label{xi}
\end{equation}

\begin{table}[!ht]
\begin{tabular}{|c|c|}
\hline \\ [-2.8ex]
Tensor & Coefficient \\
[.3ex] \hline \\ [-2.8ex]
$\partial^\mu\partial'^\nu y$ & $-(D-1)^2(2-y)f_1 +
\left[(D-1)(2-y)^2-D(4y-y^2)\right]\partial_yf_1$ \\
[.3ex] \\ [-2.8ex]
 & $+(2-y)(4y-y^2)\partial_y^2f_1-2(D-1)(2-y)\partial_uf_1$ \\
[.3ex] \\ [-2.8ex]
 & $+2\left(\frac{a}{a'}+\frac{a'}{a}\right) \left[(D-1) \partial_u
f_1 - (2-y)\partial_u \partial_y f_1\right]$ \\
[.3ex] \\ [-2.8ex]
 & $-2\left(\frac{a}{a'}-\frac{a'}{a}\right) \left[(D-1)
\partial_v f_1-(2-y)\partial_v \partial_yf_1\right]$ \\
[.3ex] \\ [-2.8ex]
 & $+2(2-y)^2\partial_u\partial_yf_1+\left[2\left(\frac{a}{a'} +
\frac{a'}{a}\right)-(2-y)\right](\partial_u^2-\partial_v^2)f_1$ \\
[.3ex] \hline \\ [-2.8ex]
$\partial^\mu y\partial'^\nu y$ & $-(D-1)^2 f_1 + (2D-1) (2-y)
\partial_yf_1-(2-y)^2\partial^2_yf_1$ \\
[.3ex] \\ [-2.8ex]
 & $-2(D-1)\partial_uf_1+2(2-y)\partial_u\partial_y f_1 -
(\partial_u^2-\partial_v^2)f_1$ \\
[.3ex] \hline \\ [-2.8ex]
$\partial^\mu y\partial'^\nu u$ & $2\frac{a'}{a}\left[-(D-1)
(\partial_u+\partial_v)f_1+(2-y)(\partial_u + \partial_v)
\partial_yf_1\right.$ \\
[.3ex] \\ [-2.8ex]
 & $\left.-(\partial_u^2-\partial_v^2)f_1\right]$ \\
[.3ex] \hline \\ [-2.8ex]
$\partial^\mu u\partial'^\nu y$ & $2\frac{a}{a'}\left[-(D-1)
(\partial_u-\partial_v)f_1+(2-y)(\partial_u - \partial_v)
\partial_yf_1\right.$ \\
[.3ex] \\ [-2.8ex]
 & $\left.-(\partial_u^2-\partial_v^2)f_1\right]$ \\
[.3ex] \hline \\ [-2.8ex]
$\partial^\mu u \partial'^\nu u$ & $-4(\partial_u^2-\partial_v^2)f_1$ \\
[.3ex] \hline
\end{tabular}

\caption{\label{f1 term} Result for expanding $D_\rho D'_\sigma
\left[(\partial^\mu\partial'^{[\nu}y)(\partial'^{\sigma]}
\partial^\rho y) f_1(y,u,v)\right]$; all coefficients are multiplied
by a factor of $\frac12 H^4$.}

\end{table}

Given the coefficient functions $F_i(y,u,v)$, one can derive
a first order differential equation in one variable for the master
structure function $\Phi(y,u,v)$ by taking a superposition of
relations (\ref{F1}-\ref{F4}),
\begin{equation}
-4 (D \!-\! 1) \Phi \!-\! 8 \partial_u \Phi = (2 \!-\! y) F_1 \!+\!
(4 y \!-\! y^2) F_2 \!+\! (2 \!-\! y) (F_3 \!+\! \widetilde{F}_3)
\!-\! F_4 \; . \qquad \label{rel1}
\end{equation}
The difference of (\ref{F3}) and (\ref{F3tilde}) gives an algebraic
relation for $\Psi(y,u,v)$ once $\Phi(y,u,v)$ is known,
\begin{equation}
-4 \sinh(v) \partial_u \Phi \!+\! 4 \cosh(v) \partial_v \Phi \!-\! 4
\sinh(v) \Psi = F_3 \!-\! \widetilde{F}_3 \; . \qquad \label{rel2}
\end{equation}
Note that de Sitter invariance implies $\Phi = \Phi(y)$ and $\Psi = 0$.

\begin{table}[!ht]
\begin{tabular}{|c|c|}
\hline \\ [-2.8ex]
Tensor & Coefficient \\
[.3ex] \hline \\ [-2.8ex]
$\partial^\mu\partial'^\nu y$ & $\left[-D(D-1)(2-y)^2+D(4y-y^2)\right]f_2-(4y-y^2)^2\partial_y^2f_2$ \\
[.3ex] \\ [-2.8ex]
 & $-(2D+1)(2-y)(4y-y^2)\partial_yf_2-2D(2-y)^2\partial_uf_2$ \\
[.3ex] \\ [-2.8ex]
 & $+2\left(\frac{a}{a'}+\frac{a'}{a}\right)\left[D(2-y)\partial_uf_2+(4y-y^2)\partial_u\partial_yf_2\right]$ \\
[.3ex] \\ [-2.8ex]
 & $-2\left(\frac{a}{a'}-\frac{a'}{a}\right)\left[D(2-y)\partial_vf_2+(4y-y^2)\partial_v\partial_yf_2\right]$ \\
[.3ex] \\ [-2.8ex]
 & $-2(2-y)(4y-y^2)\partial_u\partial_yf_2-(8-4y+y^2)(\partial_u^2-\partial^2_v)f_2$ \\
[.3ex] \\ [-2.8ex]
 & $+2(2-y)\left(\frac{a}{a'}+\frac{a'}{a}\right)(\partial^2_u-\partial_v^2)f_2$ \\
[.3ex] \hline \\ [-2.8ex]
$\partial^\mu y\partial'^\nu y$ & $-D^2(2-y)f_2+\left[-(D-1)(4y-y^2)+(D+2)(2-y)^2\right]\partial_yf_2$ \\
[.3ex] \\ [-2.8ex]
 & $+(2-y)(4y-y^2)\partial_y^2f_2-2D(2-y)\partial_uf_2$ \\
[.3ex] \\ [-2.8ex]
 & $-2(4y-y^2)\partial_u\partial_yf_2-(2-y)(\partial_u^2-\partial_v^2)f_2$ \\
[.3ex] \hline \\ [-2.8ex]
$\partial^\mu y\partial'^\nu u$ & $-2\frac{a'}{a}\left[D(2-y)(\partial_u+\partial_v)f_2+(4y-y^2)(\partial_u+\partial_v)\partial_yf_2\right]$ \\
[.3ex] \\ [-2.8ex]
 & $+\left[4-2(2-y)\frac{a'}{a}\right](\partial^2_u-\partial_v^2)f_2$ \\
[.3ex] \hline \\ [-2.8ex]
$\partial^\mu u\partial'^\nu y$ & $-2\frac{a}{a'}\left[D(2-y)(\partial_u-\partial_v)f_2+(4y-y^2)(\partial_u-\partial_v)\partial_yf_2\right]$ \\
[.3ex] \\ [-2.8ex]
 & $+\left[4-2(2-y)\frac{a}{a'}\right](\partial^2_u-\partial_v^2)f_2$\\
[.3ex] \hline \\ [-2.8ex]
$\partial^\mu u \partial'^\nu u$ & $0$ \\
[.3ex] \hline
\end{tabular}

\caption{\label{f2 term}Result for expanding $D_\rho D'_\sigma
\left[(\partial^{[\mu}y)(\partial^{\rho]}\partial'^{[\nu}y)
(\partial'^{\sigma]}y)f_2(y,u,v)\right]$; all coefficients are
 multiplied by a factor of $\frac14 H^4$.}
\end{table}

It remains to relate the master structure functions $\Phi(y,u,v)$
and $\Psi(y,u,v)$ to the structure functions $f_i(y,u,v)$ of our
representation (\ref{ansatz}-\ref{generic_tf}). Table~\ref{f1 term}
gives the coefficient functions $F_i(y,u,v)$ for the structure
function $f_1(y,u,v)$. Applying relations (\ref{rel1}-\ref{rel2})
implies that the associated master structures are,
\begin{eqnarray}
\Phi_1 & = & \frac{H^4}{2} \Bigl\{ (D \!-\! 1) f_1 \!-\! (2 \!-\! y)
\partial_y f_1 \Bigr\} \; , \label{Phi1} \\
\Psi_1 & = & \frac{H^4}{2} \Bigl\{ ( \partial_u^2 \!-\! \partial_v^2)
f_1 \Bigr\} \; . \label{Psi1}
\end{eqnarray}
Table~\ref{f2 term} gives the coefficient functions $F_i(y,u,v)$
for $f_2(y,u,v)$, from which we infer,
\begin{eqnarray}
\Phi_2 & = & \frac{H^4}{4} \Bigl\{ D (2 \!-\! y) f_2 \!+\! (4 y \!-\! y^2)
\partial_y f_2 \Bigr\} \; , \label{Phi2} \\
\Psi_2 & = & \frac{H^4}{4}\Bigl\{ (2 \!-\! y) ( \partial_u^2 \!-\!
\partial_v^2) f_2 \Bigr\} \; . \label{Psi2}
\end{eqnarray}
The master structure functions for $f_3(y,u,v)$ and
$\widetilde{f}_3(y,u,v) = f_3(y,u,-v)$ are,
\begin{eqnarray}
\Phi_3 & = & \frac{H^4}{4}  \Bigl\{ -(D \!-\! 1) (f_3 \!+\! \widetilde{f}_3)
\!+\! (2 \!-\! y)\partial_y (f_3 \!+\! \widetilde{f}_3) 
\!-\! 2  \partial_y(e^{v}
f_3 \!+\! e^{-v} \widetilde{f}_3) \Bigr\} \; , \qquad \label{Phi3} \\
\Psi_3 & = &\! \frac{H^4}{4} \Bigl\{2 \partial_y 
     \Big[\partial_u ( e^{v} f_3 \!+\! e^{-v} \widetilde{f}_3 )
        \!-\! \partial_v ( e^{v} f_3 \!-\! e^{-v} \widetilde{f}_3 )\Big]
 \!-\! (\partial_u^2 \!-\!  \partial_v^2) (f_3
\!+\! \widetilde{f}_3) \Bigr\}  . \qquad
\label{Psi3}
\end{eqnarray}
The analogous result for $f_4(y,u,v)$ is,
\begin{eqnarray}
\Phi_4 & = & \frac{H^4}{4} \Bigl\{ \partial_y f_4 \Bigr\} \; , \label{Phi4} \\
\Psi_4 & = & \frac{H^4}{4} \Bigl\{ -(D \!-\! 1) \partial_y f_4 \!+\!
(2 \!-\! y) \partial_y^2 f_4 \!-\! 2 \partial_y \partial_u f_4 \Bigr\} \; .
\qquad \label{Psi4}
\end{eqnarray}
And $f_5(y,u,v)$ gives,
\begin{eqnarray}
\Phi_5 \!\!\!\!\!& = &\!\!\!\!\! \frac{H^4}4 \Bigl\{-D f_5 \!+\! 2 (2 \!-\! y)
\partial_y f_5 \!-\! 4 \cosh(v) \partial_y f_5 \Bigr\} \; , \label{Phi5} \\
\Psi_5 \!\!\!\!\!& = &\!\!\!\!\! \frac{H^4}4 \Bigl\{\! (D \!-\! 1) f_5 \!-\!
(D \!+\! 1) (2 \!-\! y)] \partial_y f_5 \!-\! (4 y \!-\! y^2)
\partial_y^2 f_5 \!+\! 2 \partial_u f_5 
\nonumber \\
& & \hspace{0cm} 
+[-\! 2 (2 \!-\! y)\!+\! 4 \cosh(v)] \partial_u\partial_y f_5
\!-\! 4 \partial_v\partial_y [\sinh(v)f_5]
- (\partial_u^2 \!-\! \partial_v^2) f_5 
\Bigr\} 
\; . \qquad 
\label{Psi5}
\end{eqnarray}
From these expressions one sees that the vacuum polarization can be
described in terms of any two of the structure functions $f_i(y,u,v)$.
When the result is de Sitter invariant then it requires only a single
structure function, which can be either $f_1(y)$ or $f_2(y)$.

\subsection{Representation for this System}

It remains to work out the structure functions for the primitive
result (\ref{old_vacpol}). Substituting the coefficient functions
$F_i(y,u,v)$ from expression (\ref{old_vacpol}) into relation
(\ref{rel1}), and making judicious use of the $A$ propagator
equation (\ref{Aeqn}), gives the first master structure function,
\begin{eqnarray}
\Phi &=& - \frac{e^2}{2 (D \!-\!1)} \biggl[ (D \!-\! 1) [(2 \!-\! y) A'
\!-\! k] A \!+\! (4y \!-\! y^2) {A'}^2 \biggr]
\nonumber\\
&&
 -\frac{k e^2}{2} [(2 \!-\! y) A' \!-\! k] u \; ,
\label{Phi}
\end{eqnarray}
where we dropped local terms $\propto \delta^D(x-x')$
and also the prefactor $\sqrt{-g}\sqrt{-g^\prime}={\rm e}^{Du}$
in~(\ref{old_vacpol}), as it contributes to the integral measure to 
make it covariant.
Doing the same thing for relation (\ref{rel2}) implies that the
second master structure function is,
\begin{equation}
\Psi = \frac{k e^2}{2} [(2 \!-\! y) A' \!-\! k] \; .
\label{Psi}
\end{equation}

At this point we must make a choice between the ten possible pairs
of structure functions $f_i(y,u,v)$ that could be used to represent
the result. We chose $f_1$ and $f_2$, the two structure functions
that would be de Sitter invariant if the system was. In view of
relations (\ref{Psi1}), (\ref{Psi2}) and (\ref{Psi}), the requirement
that $\Psi_1 + \Psi_2 = \Psi$ implies the form,
\begin{eqnarray}
f_1(y,u,v) & = & -\frac{k^2 e^2}{4 H^4} \, (u^2 \!-\! v^2) +
f_{1a}(y,u \!+\! v) + f_{1b}(y,u \!-\! v) \; , \\
f_2(y,u,v) & = & \frac{k e^2}{2 H^4} \, A'(y) (u^2 \!-\! v^2) +
f_{2a}(y,u \!+\! v) + f_{2b}(y,u \!-\! v) \; .
\quad
\end{eqnarray}
Then requiring $\Phi_1 + \Phi_2 = \Phi$ and relations (\ref{Phi1}),
(\ref{Phi2}), and (\ref{Phi}) implies,
\begin{eqnarray}
f_1(y,u,v) & = & -\frac{k^2e^2}{4 H^4} \, (u^2-v^2) + \alpha_1(y) u
+ \alpha_2(y) \; , \\
f_2(y,u,v) & = & \frac{ke^2}{2 H^4} \, A'(y) (u^2-v^2) + \beta_1(y) u
+ \beta_2(y) \; ,
\end{eqnarray}
where the functions $\alpha_i(y)$ and $\beta_i(y)$ satisfy,
\begin{eqnarray}
&& -\frac{2ke^2}{H^4} [(2 \!-\! y) A' \!-\! k]
\nonumber \\
&&\hspace{3cm}
 =\,  2 (D \!-\! 1) \alpha_1 \!-\! 2 (2 \!-\! y) \alpha_1' \!+\!
D (2 \!-\! y) \beta_1 \!+\! (4y-y^2)\beta_1' \; , 
\label{eqn1} 
\qquad \\
&&-\frac{2 e^2}{(D \!-\! 1) H^4} \Bigl\{ (D \!-\! 1) [ (2 \!-\! y)
A' \!-\! k ] A \!+\! (4 y \!-\! y^2) {A'}^2 +\, k^2
 \Bigr\}
  \nonumber \\
& & \hspace{3cm} 
=\, 2 (D \!-\! 1) \alpha_2 \!-\! 2 (2 \!-\! y) \alpha_2' \!+\!
D (2 \!-\! y) \beta_2 \!+\! (4y-y^2)\beta_2' \; . 
\label{eqn2} \qquad
\end{eqnarray}

Because we have only two equations (\ref{eqn1}-\ref{eqn2}) in terms
of four functions $\alpha_i(y)$ and $\beta_i(y)$, there are many
solutions. Perhaps the nicest --- and one of great significance for
the next section --- results from the ansatz,
\begin{equation}
\alpha_i(y) = \gamma_i'(y) \qquad , \qquad \beta_i(y) = -\gamma_i''(y)
\; . \label{simplification}
\end{equation}
This ansatz effects the following simplification of the right hand
sides of (\ref{eqn1}-\ref{eqn2}),
\begin{eqnarray}
\lefteqn{2 (D \!-\! 1) \alpha_i(y) \!-\! 2 (2 \!-\! y) \alpha_i'(y) \!+\!
D (2 \!-\! y) \beta_i(y) \!+\! (4y-y^2)\beta_i'(y) } \nonumber \\
& & \hspace{2.4cm} = -\frac{\partial}{\partial y} \Bigl[ (4 y \!-\! y^2)
\gamma_i''(y) \!+\! D (2 \!-\! y) \gamma'_i(y) \!-\! (D \!-\! 2)
\gamma_i(y) \Bigr] \; , \qquad \\
& & \hspace{2.4cm} = -\frac{d}{d y} \Bigl[ \frac{\square}{H^2}
\!-\! (D \!-\! 2) \Bigr] \gamma_i(y) \equiv -
\frac{d}{d y} \frac{D_B}{H^2} \gamma_i(y) \; . \label{bigsimp}
\end{eqnarray}

One can see from relation (\ref{bigsimp}) that it would be highly
desirable to express the left hand sides of (\ref{eqn1}-\ref{eqn2})
as $\partial/\partial y$ of something. The desired function involves
the propagator of a scalar with mass $M^2 = (D-2) H^2$, which obeys
the equation,
\begin{equation}
D_B i\Delta_B(x;x') = \frac{i \delta^D(x \!-\! x')}{\sqrt{-g}} \; .
\label{Bpropeqn}
\end{equation}
Unlike the massless scalar, the propagator of this massive scalar
can be expressed in terms of a de Sitter invariant function of $y$,
$i\Delta_B(x;x') = B(y)$. The series expansion of this function is,
\begin{eqnarray}
\lefteqn{B(y) = \frac{H^{D-2}}{(4\pi)^{\frac{D}2}} \Biggl\{\Gamma\Bigl(
\frac{D}2 \!-\! 1\Bigr) \Bigl(\frac4{y}\Bigr)^{\frac{D}2 -1} +
\sum_{n=0}^{\infty} \Biggr[ \frac{\Gamma(n \!+\! \frac{D}2)}{\Gamma(n \!+\! 2)}
\Bigl(\frac{y}4 \Bigr)^{n -\frac{D}2 + 2} } \nonumber \\
& & \hspace{6.7cm} - \frac{\Gamma(n\!+\! D\!-\!2)}{\Gamma(n\!+\! \frac{D}2)}
\Bigl(\frac{y}4\Bigr)^{n}\Biggr]\Biggr\} \; . \qquad \label{Bexp}
\end{eqnarray}
The series expansions (\ref{Aexp}) and (\ref{Bexp}) imply two
relations between $A(y)$ and $B(y)$ of great utility,
\begin{eqnarray}
(2 \!-\! y) A'(y) \!-\! k & = & 2 B'(y) \; , \label{AtoB1} \\
(4 y \!-\! y^2) A'(y) + k (2 \!-\! y) & = & -2 (D \!-\! 2) B(y) \; .
\label{AtoB2}
\end{eqnarray}
Relation~(\ref{AtoB1}) implies that our ansatz~(\ref{simplification}) 
reduces equation (\ref{eqn1}) to the form,
\begin{equation}
D_B \gamma_1(y) = \frac{4 k e^2}{H^2} \, B(y) \; .
\end{equation}
The solution is,
\begin{equation}
\gamma_1(y) = \frac{4 k e^2}{H^2} \, i\Delta_{BB}(x;x') \; ,
\end{equation}
where the de Sitter invariant biscalar $i\Delta_{BB}(x;x')$ was
introduced in a previous study of the graviton propagator
\cite{MTW3}. Using relation (\ref{AtoB2}) similarly on (\ref{eqn2})
implies,
\begin{equation}
D_B \gamma_2(y) = 
\frac{4 e^2}{(D \!-\! 1) H^2} \int \!\! dy \Bigl\{
(D \!-\! 1) A(y) B'(y) - (D \!-\! 2) A'(y) B(y)
- k B'(y)
\Bigr\} \; .
\end{equation}
In the next section we will derive a de Sitter invariant Green's
function which can be used to solve for $\gamma_2(y)$.

\section{A Massive Scalar}\label{mneq0}

The purpose of this section is to derive a de Sitter invariant
representation for the one loop vacuum polarization from a charged
scalar with a small mass. The first subsection is devoted to
presenting the primitive result from the three diagrams in
Figs~\ref{fig1}-\ref{fig3}, with special attention to the expansion
of the massive scalar propagator for small mass. In the second
subsection we re-express the primitive result . It is at this point
that we derive the Green's function for the differential operator
$D_B$ defined in (\ref{bigsimp}). The final subsection explains
renormalization.

\subsection{The Primitive Result}

The propagator of a minimally coupled scalar of mass $M$ obeys the
equation,
\begin{equation}
(\square \!-\! M^2) i\Delta_T(x;x') = \frac{i \delta^D(x \!-\!
x')}{\sqrt{-g}} \; .
\end{equation}
For $M^2 > 0$ it has a de Sitter invariant solution $i\Delta_T(x;x')
= T(y)$,
\begin{equation}
\label{Tprop}
T(y) = \frac{H^{D-2}}{(4\pi)^{\frac{D}2}} \frac{\Gamma(\frac{D-1}{2} \!+\!
\nu) \Gamma(\frac{D-1}{2} \!-\! \nu)}{\Gamma(\frac{D}2)} \, \mbox{}_2
F_1\Bigl(\frac{D-1}{2} \!+\! \nu ; \frac{D-1}{2} \!-\! \nu ; \frac{D}2 ;
1 \!-\! \frac{y}{4}\Bigr) \; ,
\end{equation}
where $\nu\equiv\sqrt{\left(\frac{D-1}{2}\right)^2-\frac{M^2}{H^2}}$
\cite{CT}. (The $B$-type propagator (\ref{Bpropeqn}-\ref{Bexp})
represents the special case of $M^2 = (D-2) H^2$.) It is useful to
extract the most ultraviolet singular term from $T(y)$,
\begin{equation}
T(y) \equiv \frac{H^{D-2}}{4 \pi^{\frac{D}2}} \Gamma\Bigl(\frac{D}2
\!-\! 1\Bigr) \left[\frac{1}{y^{\frac{D}2 - 1}} +\Delta T(y) \right]
\equiv K \left[\frac{1}{y^{\frac{D}2 - 1}} +\Delta T(y) \right] \; .
\label{Texp}
\end{equation}

Just as in flat space, nonzero mass greatly complicates the
spacetime dependence of propagators. For example, the series
expansion of $\Delta T(y)$ in powers of $y$ does not terminate in $D
= 4$ dimensions for most choices of $M^2 > 0$. One signal that there
is no de Sitter invariant solution for the massless case is that
$\Delta T(y)$ diverges as $M$ goes to zero. This can be quantified
by giving the Laurent expansion for $\Delta T(y)$ in terms of the
parameter $s \equiv (\frac{D-1}2) - \nu$. In $D=4$ dimensions, we
have \cite{PP},
\begin{equation}
\Delta T(y) = \frac1{2 s} - \frac12 \ln\Bigl( \frac{y}4 \Bigr) - 1 
 + s \!\times\! \delta T(y) + O(s^2) \; , \label{DeltaT}
\end{equation}
where the function $\delta T(y)$ is,
\begin{equation}
\delta T(y) = \frac12{\rm Li}_2\Bigl(\frac{y}{4}\Bigr)
       +\ln\Bigl(\frac{y}{4}\Bigr)
             \biggl[\frac12\ln\Bigl(1-\frac{y}{4}\Bigr)+1-\frac{1}{4-y}\biggr]
\,,
\label{deltaT}
\end{equation}
and where ${\rm Li}_2(z)=\sum_{n=1}^\infty z^n/n^2$ is the polylogarithm 
function.
For small mass the $1/s$ contribution to the vacuum polarization is
the dominant effect.

Substituting $i\Delta_T(x;x')$ for the scalar propagator in
expression (\ref{orig_vacpol}) gives the result for the three
diagrams of Figs~\ref{fig1}-\ref{fig3}. Derivatives of the
propagator can be expressed using the tensor basis (\ref{basis})
\begin{eqnarray}
\partial_{\mu} i\Delta_T(x;x') & = & T'(y) \times \partial_{\mu} y \; , \\
\partial'_{\nu} i\Delta_T(x;x') & = & T'(y) \times \partial'_{\nu} y \; , \\
\partial_{\mu} \partial'_{\nu} i\Delta_T(x;x') & = & T'(y) \times
\partial_{\mu} \partial'_{\nu} y + T''(y) \times \partial_{\mu} y \,
\partial'_{\nu} y \nonumber \\
& & \hspace{4cm} + \frac{i \delta^D(x \!-\! x')}{H^2 \sqrt{-g} }
\times \partial_{\mu} u \, \partial'_{\nu} u \; . \qquad
\end{eqnarray}
By applying these identities we can write (\ref{orig_vacpol}) as
\begin{eqnarray}
\lefteqn{i\Bigl[\mbox{}^\mu \Pi^\nu \Bigr](x;x') = \sqrt{-g(x)} \,
g^{\mu\rho}(x) \sqrt{-g(x')} \, g^{\nu\sigma}(x') \biggl\{
\partial_\rho \partial'_\sigma y \!\times\! (-2 e^2 T T') }
\nonumber \\
& & \hspace{.5cm} + \partial_\rho y\partial'_\sigma y \!\times\! 2
e^2 \Bigl({T'}^2 \!-\! T T''\Bigr) \!-\! 2 e^2 T(0) \Bigl[
g_{\rho\sigma} \!+\! H^{-2} \partial_{\rho} u \partial_{\sigma}
u\Bigr] \frac{i \delta^D(x \!-\! x')}{\sqrt{-g}} \nonumber \\
& & \hspace{3cm} + \delta Z \biggl[ g_{\rho\sigma} \Bigl[ \square
\!+\! (D \!-\! 1) H^2 \Bigr] \!-\! D_{\rho} D_{\sigma} \biggr]
\frac{i \delta^D(x-x')}{\sqrt{-g}} \biggr\} \; , \qquad
\label{old_vacpol2}
\end{eqnarray}
where 
\begin{equation}
T(0) = \frac{H^{D-2}}{(4\pi)^{\frac{D}2}} 
\frac{\Gamma(\frac{D-1}{2} \!+\!\nu) \Gamma(\frac{D-1}{2} \!-\! \nu)}
     {\Gamma(\frac{1}2+\nu)\Gamma(\frac{1}2-\nu)} 
\,.
\nonumber
\end{equation}

\subsection{The $\gamma(y)$ Representation}

Because the primitive result (\ref{old_vacpol2}) is de Sitter
invariant, it can be given a manifestly de Sitter invariant
expression using the structure functions $f_1(y)$ and $f_2(y)$ of
our general representation (\ref{ansatz}-\ref{generic_tf}). A
further simplification arises if we relate the structure functions
as $f_1(y) = \gamma'(y)$ and $f_2(y) = -\gamma''(y)$,
\begin{eqnarray}
\lefteqn{i \Bigl[ \mbox{}^{\mu} \Pi^{\nu} \Bigr](x;x') =
\sqrt{-g(x)} \, g^{\mu\rho}(x) \sqrt{-g(x')} \, g^{\nu\sigma}(x')
\times D^{\alpha} {D'}^{\beta} } \nonumber \\
& & \hspace{2.5cm} \times \Biggl\{ \partial_{\rho}
\partial'_{[\sigma} y \, \partial'_{\beta]} \partial_{\alpha} y
\!\times\! \gamma'(y) \!-\! \partial_{[\rho} y \, \partial_{\alpha]}
\partial'_{[\sigma} y \, \partial'_{\beta]} y \!\times\! \gamma''(y)
\Biggr\} \; , \qquad \\
& & \hspace{1.8cm} = \sqrt{-g} \, g^{\mu\rho} \sqrt{-g'} \,
{g'}^{\nu\sigma} \times D^{\alpha} {D'}^{\beta} D_{[\rho}
D'_{[[\sigma} \Bigl\{ \partial_{\alpha]} \partial'_{\beta]]} y
\!\times\! \gamma(y) \Bigr\} \; . \qquad \label{gammarep}
\end{eqnarray}
The double brackets used here and henceforth serve to distinguish
which index group is anti-symmetrized,
\begin{eqnarray}
\lefteqn{D_{[\rho} D'_{[[\sigma} \Bigl\{ \partial_{\alpha]}
\partial'_{\beta]]} y \!\times\! \gamma \Bigr\} \equiv
\frac14 D_{\rho} D'_{\sigma} \Bigl\{ \partial_{\alpha}
\partial'_{\beta} y \!\times\! \gamma \Bigr\} -
\frac14 D_{\rho} D'_{\beta} \Bigl\{ \partial_{\alpha}
\partial'_{\sigma} y \!\times\! \gamma \Bigr\} } \nonumber \\
& & \hspace{3.5cm} -\frac14 D_{\alpha} D'_{\sigma} \Bigl\{
\partial_{\rho} \partial'_{\beta} y \!\times\! \gamma \Bigr\}
+\frac14 D_{\alpha} D'_{\beta} \Bigl\{ \partial_{\rho}
\partial'_{\sigma} y \!\times\! \gamma \Bigr\} \; . \qquad
\end{eqnarray}

The relations $f_1(y) = \gamma'(y)$ and $f_2(y) = -\gamma''(y)$ are
the same as (\ref{simplification}) that was considered in the
previous section. A simple consequence of applying that analysis to
(\ref{old_vacpol2}), without worrying about the delta function
contributions, is the relation,
\begin{equation}
(D \!-\! 1) H^4 \frac{\partial}{\partial y} \Bigl[ \frac{D_B}{H^2}
\, \gamma(y) \Bigr] = 2 e^2 \Bigl\{ (4 y \!-\! y^2) ( {T'}^2 \!-\! T
T'') \!-\! (2 \!-\! y) T T' \Bigr\} \; ,
\end{equation}
where $D_B \equiv \square - (D-2) H^2$. If we introduce the
indefinite integral symbol, $I[f] \equiv \int dy f(y)$, the
differential equation for $\gamma(y)$ becomes,
\begin{equation}
\frac{D_B}{H^2} \, \gamma(y) = \frac{2 e^2}{(D \!-\! 1) H^4} \,
I\Bigl[ (4 y \!-\! y^2) ( {T'}^2 \!-\! T T'') \!-\! (2 \!-\! y) T T'
\Bigr] \equiv \mathcal{S}(y) \; . \label{gammaeqn}
\end{equation}
It turns out that enforcing this equation automatically recovers the
undifferentiated delta functions of expression (\ref{old_vacpol2}).
We will shortly demonstrate that a special choice for the homogeneous
part of the solution recovers the contribution from the field strength
renormalization.

If we ignore delta function contributions, the action of $D_B$ on a
function of $y$ can be written as,
\begin{equation}
\frac{D_B}{H^2} \, \gamma(y) = (4 y \!-\! y^2) \gamma''(y) + D (2
\!-\! y) \gamma'(y) - (D \!-\! 2) \gamma(y) \; .
\end{equation}
This is a second order, linear differential operator in $y$ whose
two homogeneous solutions are easily seen to be the $B$-type
propagator and its translation to the antipodal point,
\begin{equation}
\gamma_1(y) = B(y) \qquad , \qquad \gamma_2(y) = B(4 \!-\! y) \; .
\end{equation}
From (\ref{Bexp}) we find that their Wronskian is,
\begin{equation}
W(y) \equiv \gamma_1(y) \gamma_2'(y) \!-\! \gamma_1'(y) \gamma_2(y)
= \frac{H^{2 D - 4}}{4 \pi^D} \frac{\Gamma(\frac{D}2)
\Gamma(\frac{D}2 \!-\! 1)}{ (4 y \!-\! y^2)^{\frac{D}2}} \; .
\end{equation}
Of course acting $D_B$ on $\gamma_1(y)$ really produces a delta
function at $y=0$, just as acting $D_B$ on $\gamma_2(y)$ produces a
delta function at the antipodal point $y = 4$. The unique Green's
function which avoids both poles is,
\begin{equation}
G_B(y;y') = -\frac{ \theta(y \!-\! y') \gamma_1(y) \gamma_2(y')
\!+\! \theta(y' \!-\! y) \gamma_2(y) \gamma_1(y') }{ (4 y' \!-\!
{y'}^2) W(y') } \; . \label{Gfunct}
\end{equation}

Possession of a Green's function such as (\ref{Gfunct}) immediately
defines the solution of (\ref{gammaeqn}) up to homogeneous
contributions. Expression (\ref{old_vacpol2}) contains no delta
functions which become singular at the antipodal point, so there can
be no contamination from $B(4-y)$. However, (\ref{old_vacpol2}) does
contain delta function which become singular at $y=0$, so we must
allow a term proportional to $B(y)$. Hence the solution for
$\gamma(y)$ takes the form,
\begin{equation}
\gamma(y) = {\rm Const} \!\times\! B(y) + \int_{0}^{4} \!\! dy' \,
G_B(y;y') \, \mathcal{S}(y') \; . \label{gammaform}
\end{equation}

To fix the homogeneous term we act the four derivatives of
(\ref{gammarep}) on $B(y)$,
\begin{eqnarray}
\label{B} \lefteqn{ D^{\alpha} {D'}^{\beta} D_{[\rho} D'_{[[\sigma}
\Bigl\{ \partial_{\alpha]} \partial'_{\beta]]} y \!\times\!
B(y)\Bigr\} = -\frac14 H^4 (D \!-\! 1)^2 (2 \!-\! y) \partial_{\rho}
\partial'_{\sigma} B } \nonumber \\
& & \hspace{.5cm} -H^2 (D\!-\!1) \Bigl[\partial^{\alpha} y \,
D_{\alpha} \!+\! {\partial'}^{\beta}y \, D'_{\beta} \Bigr]
\partial_{\rho} \partial'_{\sigma} B + \partial^{\alpha}
{\partial'}^{\beta} y \!\times\! D_{\alpha} D'_{\beta}
\partial_{\rho} \partial'_{\sigma} B \nonumber \\
& & \hspace{1.5cm} + H^2 (D \!-\! 1) \Bigl[\partial_{\rho} y
\!\times\! \square \partial'_{\sigma} B \!+\! \partial'_{\sigma} y
\!\times\! \square' \partial_{\rho} B\Bigr] - \partial_{\rho}
\partial'_{\beta} y \!\times\! \square {D'}^{\beta}
\partial'_{\sigma} B \nonumber \\
& & \hspace{5cm} -\partial_{\alpha} \partial'_{\sigma} y \!\times\!
\square' D^{\alpha} \partial_{\rho} B \!+\! \partial_{\rho}
\partial'_{\sigma} y \!\times\! \square \square' B \; . \qquad
\end{eqnarray}
Of course we already know that expression (\ref{B}) must vanish
except for delta function terms. We can collect the various delta
function terms by making use of the identities,
\begin{eqnarray}
\square B & = & \square' B = (D \!-\! 2) H^2 B \!+\!
\frac{i\delta^D(x-x')}{\sqrt{-g}} \; , \\
D_{\alpha} D'_{\beta} B & = & \partial_{\alpha} \partial'_{\beta} y
\!\times\! B' \!+\! \partial_{\alpha} y \, \partial'_{\beta} y
\!\times\! B'' \!+\! \partial_{\alpha} u \, \partial'_{\beta} u
\!\times\! \frac{i\delta^D(x-x')}{\sqrt{-g}} \; , \\
\frac{\square}{H^2}I[B] & = & (4y \!-\! y^2) B' \!+\! D(2 \!-\! y)B
= 2(2 \!-\! y) B \!-\! 2k \; .
\end{eqnarray}
After much work we find,
\begin{eqnarray}
\label{counterterm} \lefteqn{D^{\alpha} {D'}^{\beta} D_{[\rho}
D'_{[[\sigma} \Bigl\{ \partial_{\alpha]} \partial'_{\beta]]} y
\!\times\! B(y) \Bigr\} } \nonumber \\
& & \hspace{2.5cm} = - \frac12 H^2 \Biggl[ g_{\rho\sigma} \Bigl[
\square \!+\! (D \!-\! 1) H^2 \Bigr] \!-\! D_{\rho} D'_{\sigma}
\Biggr] \frac{i\delta^D(x-x')}{\sqrt{-g}} \; . \qquad
\end{eqnarray}
Comparison with (\ref{old_vacpol2}) reveals that the constant in
expression (\ref{gammaform}) must be $-2 \delta Z/H^2$,
\begin{equation}
\gamma(y) = -\frac{2 \delta Z}{H^2} \!\times\! B(y) + \int_{0}^{4}
\!\! dy' \, G_B(y;y') \, \mathcal{S}(y') \; . \label{gammafinal}
\end{equation}

\subsection{Renormalization}

It is both tedious and unnecessary to maintain full dimensional
regularization when integrating the Green's function up against the
source in expression (\ref{gammafinal}). The primitive result
(\ref{old_vacpol2}) is only quadratically divergent, which means it
goes like $1/y^3$ near coincidence in $D=4$ dimensions. Hence
extracting four derivatives, the way we do in (\ref{gammarep}),
leaves the most singular term in $\gamma(y)$ behaving like $1/y$
near coincidence. This is integrable, so we could set $D=4$
directly, except for the fact that the $D$-dependent coefficient of
this leading term happens to diverge like $1/(D-4)$. It is only
necessary to keep this leading term in $D$ dimensions; the remainder
can be evaluated in $D=4$ dimensions. Of course the divergence is
absorbed by the field strength renormalization. We work out the
details below.

The analysis begins by employing expressions (\ref{Texp}-\ref{deltaT})
to infer,
\begin{eqnarray}
\lefteqn{(4 y \!-\! y^2) ({T'}^2 \!-\! T T'') \!-\! (2 \!-\! y) T T' }
\nonumber \\
& & \hspace{0cm}  = K^2 \Biggl\{ -\frac{(D \!-\! 2)}{y^{D-1}} \!-\!
\frac{3}{s y^2} \!+\! \frac{3 \ln(\frac{y}4)}{y^2} \!+\! \frac{9}{y^2}
\!+\!  \frac{\frac34}{y} \!+\! \frac{3 \ln(\frac{y}4)}{(4 \!-\! y)^2}
\!+\! \frac{\frac34}{4 \!-\! y} + O(s) \Biggr\} \; , \qquad
\label{doubleT}
\end{eqnarray}
where we recall that $s \equiv (\frac{D-1}2) - \nu$ vanishes as the
scalar mass goes to zero. Now substitute (\ref{doubleT}) in the
definition (\ref{gammaeqn}) of the source to obtain,
\begin{eqnarray}
\mathcal{S}(y) & \equiv & \frac{2 e^2}{(D \!-\! 1) H^4} \,
I\Bigl[ (4 y \!-\! y^2) ({T'}^2 \!-\! T T'') \!-\! (2 \!-\! y)
T T' \Bigr] \; , \qquad \\
& = & \frac{2 e^2 K^2}{(D \!-\! 1) H^4} \Biggl\{ \frac1{y^{D-2}} \!+\!
\frac{3}{s y} \!-\! \frac{3 \ln(\frac{y}4)}{y} \!-\! \frac{12}{y}
\!+\! \frac{3 \ln(\frac{y}4)}{4 \!-\! y} + O(s) \Biggr\} \; , \qquad \\
& \equiv & \frac{2 e^2 K^2}{(D \!-\! 1) H^4} \times \frac1{y^{D-2}} +
\mathcal{S}_R(y) \; . \label{rhs}
\end{eqnarray}
This distinguishes the most singular part of the source from the less
singular remainder, $\mathcal{S}_R(y)$.

The next step is to make a similar distinction between the most
singular part of $\gamma(y)$ and the less singular remainder
$\gamma_R(y)$,
\begin{equation}
\gamma(y) \equiv -\frac{2 \delta Z}{H^2} \times B(y) +
\frac{\alpha}{y^{D-3}} + \gamma_R(y) \; .
\end{equation}
If we do not worry about delta functions, the result of acting
$D_B/H^2$ is,
\begin{equation}
\frac{D_B}{H^2} \, \gamma(y) = \frac{2 (D \!-\! 3) (D \!-\! 4)
\alpha}{y^{D-2}} + \frac{ (D \!-\! 4) \alpha}{ y^{D-3}} +
\frac{D_B}{H^2} \, \gamma_{R}(y) \; . \label{lhs}
\end{equation}
Equating (\ref{lhs}) to (\ref{rhs}) implies that the coefficient
$\alpha$ is,
\begin{equation}
\alpha = \frac{e^2 K^2}{(D \!-\! 1) (D \!-\! 3) (D \!-\! 4) H^4} \; ,
\end{equation}
and also that (for $D=4$) the residual part of $\gamma$ obeys,
\begin{equation}
\frac{D_B}{H^2} \, \gamma_R(y) = \mathcal{S}_R(y) - \frac{e^2}{48
\pi^4} \!\times\! \frac1{y} = \frac{e^2}{8 \pi^4} \Biggl\{
-\frac{\ln(\frac{y}4)}{y} + \frac{[\frac1{s} \!-\! \frac{25}6]}{y} +
\frac{\ln(\frac{y}4)}{4 \!-\! y} + O(s) \!\Biggr\} . \label{remeqn}
\end{equation}

At this stage we can combine the potentially divergent parts of
$\gamma(y)$,
\begin{equation}
\gamma(y) - \gamma_R(y) = -\frac{2 \delta Z}{H^2} \, B(y) +
\frac{e^2 K^2}{(D \!-\! 1)(D \!-\! 3) (D\!-\! 4) H^4} \,
\frac1{y^{D-3}} \; . \label{potsing}
\end{equation}
Now recall from (\ref{Bexp}) that $B(y) = K/y^{\frac{D}2-1} + O(D-4)$.
The divergent parts of (\ref{potsing}) will cancel in $D=4$ dimensions
if we choose,
\begin{equation}
\delta Z = \frac{e^2 K}{2 (D \!-\! 1) (D \!-\! 3) (D \!-\! 4) H^2} \; .
\label{deltaZ}
\end{equation}
This permits us to finally take the unregulated limit,
\begin{equation}
\lim_{D \rightarrow 4} \Bigl[\gamma(y) - \gamma_R(y) \Bigr] =
\frac{e^2}{96 \pi^4} \Bigl\{ - \frac{\ln(y)}{y} + \frac{
\ln(\frac{y}4)}{4 \!-\! y} \Bigr\} \; . \label{gammalim}
\end{equation}
We note in passing that the divergent part of the field strength
renormalization (\ref{deltaZ}) agrees with the two previous de Sitter
results \cite{PTW,PP} and, indeed, with the flat space result.

It remains to evaluate $\gamma_R(y)$ in $D=4$ dimensions. From
equations (\ref{remeqn}) and (\ref{Gfunct}) we have,
\begin{eqnarray}
\lefteqn{\gamma_R(y) = -\frac1{4 y} \int_0^y \!\! dy' \, y' \Bigl[
\mathcal{S}_R(y') \!-\! \frac{e^2}{48 \pi^4} \, \frac1{y'} \Bigr] }
\nonumber \\
& & \hspace{4cm} - \frac1{4 (4 \!-\! y)} \int_y^4 \!\! dy' \, (4
\!-\! y') \Bigl[ \mathcal{S}_R(y') \!-\! \frac{e^2}{48 \pi^4} \,
\frac1{y'} \Bigr] \; , \qquad \\
& & \hspace{.5cm} = \frac{e^2}{8 \pi^4} \Biggl\{ -\frac{
\ln^2(\frac{y}{4})}{2 (4 \!-\! y)} + \Bigl[ \frac1{s} \!-\!
\frac{13}6\Bigr] \frac{ \ln(\frac{y}4)}{4 \!-\! y} + \frac{[
\frac{\pi^2}{6} \!-\! {\rm Li}_2(1 \!-\! \frac{y}4)]}{y}
+ O(s) \Biggr\} \; . \qquad \label{gammaR}
\end{eqnarray}
Combining (\ref{gammalim}) and (\ref{gammaR}) gives us the fully
renormalized structure function,
\begin{equation}
\gamma_{\rm ren} = \frac{e^2}{8 \pi^4} \Biggl\{\!- \frac{ \ln(y)}{12
y} -\frac{ \ln^2(\frac{y}{4})}{2 (4 \!-\! y)} +\! \Bigl[ \frac1{s} -
\frac{25}{12} \Bigr] \frac{ \ln(\frac{y}4)}{4 \!-\! y} + \frac{[
\frac{\pi^2}{6} \!-\! {\rm Li}_2(1 \!-\! \frac{y}4)]}{y} + O(s)
\!\Biggr\} . \label{gammaren}
\end{equation}
Here ${\rm Li}_2(z) \equiv -\int_0^z dt \ln(1-t)/t$ is the
Dilogarithm Integral.

\section{Using the de Sitter Invariant Form}\label{dsinv}

The purpose of this section is to demonstrate that the de Sitter
invariant representation derived in the previous section does not
provide a transparent expression for the effective field equations.
In the first subsection we partially integrate the Schwinger-Keldysh
effective field equations to obtain the startling result that all
quantum corrections for dynamical photons can be expressed as
surface terms at the initial time. Such surface terms are usually
suppressed by inverse powers of the scale factor, and ignoring them
enormously simplifies the analysis. To show that this is not
possible with the cumbersome, de Sitter invariant formulation, the
second subsection focusses on the contribution which diverges when
the scalar mass goes to zero. In the vastly simpler, noncovariant
representation this contribution takes the form of a local photon
mass \cite{PP}. We at length reach the same form using Green's 2nd
Identity.

\subsection{A Paradox from the Effective Field Equations}

The quantum-corrected effective field equations are,
\begin{equation}
\partial_{\nu} \Bigl[ \sqrt{-g} \, g^{\nu\rho} g^{\mu\sigma}
F_{\rho\sigma}(x) \Bigr] + \int \!\! d^4x' \Bigl[ \mbox{}^{\mu}
\Pi^{\nu}\Bigr](x;x') A_{\nu}(x') = J^{\mu}(x) \; . \label{QMax}
\end{equation}
Here the vacuum polarization takes the form (\ref{gammarep}),
\begin{equation}
\Bigl[\mbox{}^{\mu} \Pi^{\nu} \Bigr](x;x') = -i \sqrt{-g} \sqrt{-g'}
D_{\rho} {D'}_{\sigma} D^{[\mu} {D'}^{[[\nu} \Bigl\{ D^{\rho]}
{D'}^{\sigma]]} y \!\times\! \gamma_{SK} \Bigr\} \; . \label{ourPi}
\end{equation}
However, the structure function $\gamma_{SK}$ is not quite the
in-out structure function (\ref{gammaren}) that was derived in the
previous section. If the effective field $A_{\mu}(x)$ is to
represent the true expectation value --- rather than the in-out
matrix element --- of the vector potential operator in the presence
of some state (released at some finite time $\eta_i$) then
$\gamma_{SK}$ must be the Schwinger-Keldysh structure function
\cite{SK},
\begin{equation}
\gamma_{SK} = \gamma_{\scriptscriptstyle ++} +
\gamma_{\scriptscriptstyle +-} \; .
\end{equation}
At the one loop order we are working, $\gamma_{\scriptscriptstyle
++}$ is expression (\ref{gammaren}) with the replacement \cite{FW},
\begin{equation}
y(x;x') \longrightarrow y_{\scriptscriptstyle ++}(x;x') \equiv a a'
H^2 \Bigl[ \Bigl\Vert \vec{x} \!-\! \vec{x}' \Bigr\Vert^2 - \Bigl(
|\eta \!-\! \eta'| \!-\! i \delta \Bigr)^2 \Bigr] \; . \label{y++}
\end{equation}
At the same order, $\gamma_{\scriptscriptstyle +-}$ is just minus
(\ref{gammaren}) with the replacement \cite{FW},
\begin{equation}
y(x;x') \longrightarrow y_{\scriptscriptstyle +-}(x;x') \equiv a a'
H^2 \Bigl[ \Bigl\Vert \vec{x} \!-\! \vec{x}' \Bigr\Vert^2 - \Bigl(
\eta \!-\! \eta' \!+\! i \delta \Bigr)^2 \Bigr] \; . \label{y+-}
\end{equation}
And the spacetime integration in (\ref{QMax}) is \cite{FW},
\begin{equation}
\int \!\! d^4x' = \int_{\eta_i}^0 \!\! d\eta' \! \int \!\! d^3x'
\end{equation}

For $\eta' > \eta$ we see from expressions (\ref{y++}) and
(\ref{y+-}) that $y_{\scriptscriptstyle ++}(x;x')$ equals
$y_{\scriptscriptstyle +-}(x;x')$, which means
$\gamma_{\scriptscriptstyle ++}$ cancels $\gamma_{\scriptscriptstyle
+-}$ and $\gamma_{SK}$ vanishes. There is a similar cancellation for
${x'}^{\mu}$ outside the past light-cone of $x^{\mu}$. Inside the
past light-cone we have $\gamma_{\scriptscriptstyle +-} =
-\gamma_{\scriptscriptstyle ++}^*$, which cancels the factor of $i$
in (\ref{ourPi}). Therefore, the Schwinger-Keldysh formalism is both
real and causal.

We come now to the problem with the de Sitter invariant
representation (\ref{ourPi}), which is what to do with the four
external derivatives. It is natural to extract the two derivatives
with respect to $x^{\mu}$ from inside the integral,
\begin{eqnarray}
\lefteqn{\int \!\! d^4x' \, \Bigl[\mbox{}^{\mu} \Pi^{\nu}
\Bigr](x;x') A_{\nu}(x') } \nonumber \\
& & = -i \sqrt{-g} \, D_{\rho} D^{[\mu} \!\! \int \!\! d^4x'
\sqrt{-g'} \, {D'}_{\sigma} {D'}^{[[\nu} \Bigl\{ D^{\rho]}
{D'}^{\sigma]]} y \!\times\! \gamma_{SK} \Bigr\} \!\times\!
A_{\nu}(x') \; . \qquad \label{goodstep}
\end{eqnarray}
In fact, this is {\it mandatory} if we are to take the unregulated
limit. It is just as natural --- and just as mandatory --- to
partially integrate the two derivatives with respect to
${x'}^{\mu}$,\footnote{We normalize the scale factor to unity at
$\eta = \eta_i$ so that $\sqrt{-g(\eta_i,\vec{x})} = 1$.}
\begin{eqnarray}
\lefteqn{\int \!\! d^4x'\, \Bigl[ \mbox{}^{\mu} \Pi^{\nu} \Bigr](x;x')
\, A_\nu(x')} \nonumber \\
& & = \frac{i}2 \sqrt{-g} \, D_\rho D^{[\mu} \!\! \int_{\eta_i}
\!\!\! d^3x' \Bigl[ D^{\rho]} D'^0 y \!\times\! {D'}^{\nu}
\gamma_{SK} \!-\! D^{\rho]} {D'}^{\nu} y \!\times\! {D'}^{0}
\gamma_{SK} \Bigr] A_{\nu}(x') \nonumber \\
& & \hspace{1.5cm} - \frac{i}2 \sqrt{-g} \, D_\rho D^{[\mu} \!\!
\int_{\eta_i} \!\!\! d^3x' D^{\rho]} D'_{\sigma} y \!\times\!
\gamma_{SK} F^{\sigma 0}(x') \nonumber \\
& & \hspace{2.5cm} +\frac{i}2 \sqrt{-g} \, D_\rho D^{[\mu} \!\! \int
\!\! d^4x' \! \sqrt{-g'} \, D^{\rho]} {D'}_{\sigma} y \!\times\!
\gamma_{SK} {D'}_{\nu} F^{\nu\sigma}(x') \; . \qquad \label{qcor}
\end{eqnarray}
Note that the only possible surface terms in the Schwinger-Keldysh
formalism are at the initial time; $\gamma_{\scriptscriptstyle +-}$
cancels $\gamma_{\scriptscriptstyle ++}$ at the future time surface
and at spatial infinity.

Integrals over the initial value surface are ubiquitous in
computations using the Schwinger-Keldysh formalism
\cite{PTW,PP,PW,PTsW,KW,OW,FW,PW2,BOW,DW,MW,KW2,SPW1,SPW2,LW},
and much has been learned from all this experience. One typical
(but not universal) feature of such surface terms is that they
fall off like powers of the scale factor, which rapidly makes
them irrelevant. Another feature is that they are liable to be
absorbed by perturbative corrections of the initial state
\cite{OW,FW}. The required state correction has been explicitly
constructed in one case \cite{KOW}.

Because contributions from the initial value surface tend to
redshift rap\-id\-ly, and can sometimes be absorbed into
perturbative corrections of the initial state, it has become common
to ignore them \cite{PTsW,SPW2,LW}. This vastly simplifies
computations, {\it but it cannot be possible for the surface
integrals in expression (\ref{qcor}).} To see why, we substitute
(\ref{qcor}) in the quantum-corrected Maxwell equation
({\ref{QMax}), without the surface integrals, and specialize to the
case of dynamical photons with $J^{\mu}(x) = 0$. For simplicity, we
also delete the overall factor of $\sqrt{-g}$,
\begin{equation}
D_{\nu} F^{\nu\mu}(x) + \frac{i}2 D_{\rho} D^{[\mu} \!\! \int \!\!
d^4x' \sqrt{-g'} \, D^{\rho]} D'_{\sigma} y \!\times\! \gamma_{SK}
\!\times\! D'_{\nu} F^{\nu \sigma}(x') = 0 \label{badeqn}
\end{equation}
It is immediately obvious that any field which obeys the classical
equation $D_{\nu} F^{\nu\mu} = 0$ also solves (\ref{badeqn}). One
might worry about the possibility of additional solutions resulting
from cancellations between the local and nonlocal terms in (\ref{badeqn}).
However, these sorts of solutions --- if they even exist --- are
never valid within the perturbative framework imposed by only
possessing the vacuum polarization to some finite order in the
loop expansion \cite{JZS}. We therefore conclude that dynamical
photons would receive no corrections, to any order, if it is valid
to ignore surface integrals. The problem with this conclusion is
that the vastly simpler, noncovariant formalism allows one to prove
that quantum corrections for this model change dynamical photons
from massless to massive \cite{PP}.

\subsection{Resolution using Green's 2nd Identity}

When a single assumption leads to false results, that assumption can
not be correct. In this subsection we demonstrate that the surface
integrals in expression (\ref{qcor}) cannot be dropped. To simplify
the analysis we focus on the $1/s$ contribution to $\gamma_{SK}$,
which should dominate for the case of small scalar mass,
\begin{equation}
\gamma_{SK} \longrightarrow \gamma_s \equiv \frac{e^2}{8 \pi^4 s}
\Biggl\{ \frac{\ln(\frac{y_{\scriptscriptstyle ++}}4)}{4 \!-\!
y_{\scriptscriptstyle ++}} - \frac{\ln(\frac{y_{\scriptscriptstyle
+-}}4)}{4 \!-\! y_{\scriptscriptstyle +-}} \Biggr\} \; .
\label{gammas}
\end{equation}
For any transverse vector potential ($D^{\nu} A_{\nu} = 0$) the
noncovariant formalism implies that the $1/s$ part of the full
vacuum polarization reduces to a local photon mass term \cite{PP},
\begin{equation}
\int \!\! d^4x' \Bigl[ \mbox{}^{\mu} \Pi^{\nu}\Bigr](x;x')
A_{\nu}(x') = -\frac{e^2 H^2}{4 \pi^4 s} \sqrt{-g(x)} g^{\mu\nu}(x)
A_{\nu}(x) + O(s^0) \; . \label{Proca}
\end{equation}
It must therefore be that $\gamma_s$ gives the same result in the de
Sitter invariant formalism.

Our starting point is the observation that the right hand side of
expression (\ref{qcor}) bears a close relation to Green's Second
Identity. To see this, let us define the vector ``pseudo-Green's
function'',
\begin{equation}
\Bigl[ \mbox{}^{\mu} G^{\nu} \Bigr](x;x') \equiv D_{\rho} D^{[\mu}
\Bigl[ D^{\rho]} {D'}^{\nu} y \!\times\! \frac{i}2 \, \gamma_{SK}
\Bigr] \; . \label{pseudoGreen}
\end{equation}
For any transverse vector potential expression (\ref{qcor}) can be
written,
\begin{eqnarray}
\lefteqn{ \frac1{\sqrt{-g}} \! \int\!\! d^4x' \Bigl[ \mbox{}^{\mu}
\Pi^{\nu}\Bigr](x;x') A_{\nu}(x') = \! \int \!\! d^4x' \Bigl[
\mbox{}^{\mu} G^{\nu}\Bigr](x;x') \Bigl[ \square' \!-\! (D \!-\! 1)
H^2\Bigr] A_{\nu}(x') } \nonumber \\
& & \hspace{0cm} - \int_{\eta_i} \!\!\! d^3x' \Biggl\{ {D'}^0 \Bigl[
\mbox{}^{\mu} G^{\nu}\Bigr](x;x') \!\times\! A_{\nu}(x') - \Bigl[
\mbox{}^{\mu} G^{\nu} \Bigr](x;x') \!\times\! {D'}^0 A_{\nu}(x')
\Biggr\} \; . \qquad \label{our2ndID}
\end{eqnarray}
Of course the right hand side of (\ref{our2ndID}) is what one gets
by integrating Green's Second Identity,
\begin{eqnarray}
\lefteqn{ \Bigl[ \square' \!-\! (D \!-\! 1) H^2\Bigr] \Bigl[
\mbox{}^{\mu} G^{\nu}\Bigr](x;x') \!\times\! A_{\nu}(x') } \nonumber
\\
& & \hspace{1cm} = \Bigl[ \mbox{}^{\mu} G^{\nu}\Bigr](x;x')
\!\times\! \Bigl[ \square' \!-\! (D \!-\! 1) H^2\Bigr] A_{\nu}(x')
\nonumber \\
& & \hspace{1cm} + D'_{\rho} \Biggl\{ {D'}^{\rho} \Bigl[
\mbox{}^{\mu} G^{\nu}\Bigr](x;x') \!\times\! A_{\nu}(x') \!-\!
\Bigl[ \mbox{}^{\mu} G^{\nu}\Bigr](x;x') \!\times\! {D'}^{\rho}
A_{\nu}(x') \Biggr\} \; . \qquad \label{true2ndID}
\end{eqnarray}
Relation (\ref{Proca}) will be established if we can show that
replacing $\gamma_{SK}$ with just $\gamma_s$ in (\ref{pseudoGreen})
gives a true Green's function (for transverse vectors) with
coefficient $-e^2 H^2/(4 \pi^2 s)$.

The initial part of the analysis does not require the precise form
of the structure function, so we consider a general function $f(y)$,
\begin{eqnarray}
\lefteqn{ D_{\rho} D^{[\mu} \Bigl[ D^{\rho ]} {D'}^{\nu} y
\!\times\! f \Bigr] = \frac12 H^2 \Biggl\{ D^{\mu} {D'}^{\nu} y
\Bigl[ -(4 y \!-\! y^2) f'' \!-\! (D \!-\! 1) (2 \!-\! y) f'
\Bigr] } \nonumber \\
& & \hspace{5cm} + D^{\mu} y \, {D'}^{\nu} y \Bigl[ (2 \!-\! y) f''
\!-\! (D \!-\! 1) f' \Bigr] \Biggr\} \; . \qquad \label{step1}
\end{eqnarray}
The next step is to act the photon kinetic operator, which we shall
do ignoring potential delta function contributions,
\begin{eqnarray}
\lefteqn{ \Bigl[ \square \!-\! (D \!-\! 1) H^2 \Bigr] D_{\rho}
D^{[\mu} \Bigl[ D^{\rho ]} {D'}^{\nu} y \!\times\! f \Bigr] =
\Bigl( {\rm delta\ functions}\Bigr) } \nonumber \\
& & \hspace{.5cm} + \frac12 H^4 \Biggl\{ D^{\mu} {D'}^{\nu} y \Bigl[
-(4 y \!-\! y^2) F'' \!-\! (D \!-\! 1) (2 \!-\! y) F' \Bigr]
\nonumber \\
& & \hspace{5cm} + D^{\mu} y \, {D'}^{\nu} y \Bigl[ (2 \!-\! y) F''
\!-\! (D \!-\! 1) F' \Bigr] \Biggr\} \; . \qquad \label{step2}
\end{eqnarray}
Here the function $F(y)$ is,
\begin{equation}
F(y) \equiv (4 y \!-\! y^2) f''(y) \!+\! D (2 \!-\! y) f'(y) \!-\!
(D \!-\! 2) f(y) \equiv \frac{D_B}{H^2} f(y) \; .
\end{equation}

At this stage we specialize to $f(y) = \frac{i}2 \gamma_{s}$, and
also set $D=4$. Because $B(y) = H^2/4\pi^2 y$ one has,
\begin{equation}
\frac{D_B}{H^2} \Bigl[ \frac{i}2 \gamma_s \Bigr] = \frac{i e^2}{4
\pi^2 s H^2} \Bigl[ B( y_{\scriptscriptstyle ++}) \!-\!
B(y_{\scriptscriptstyle +-}) \Bigr] \; .
\end{equation}
Now use relations (\ref{AtoB1}-\ref{AtoB2}) and (\ref{Aeqn}) to
infer,
\begin{eqnarray}
-(4 y^2 \!-\! y) B''(y) - (D \!-\! 1) (2 \!-\! y) B'(y) & = & 2
A'(y) \; , \label{AID1} \\
(2 \!-\! y) B'(y) - (D \!- \! 1) B(y) & = & 2 A''(y) \; .
\label{AID2}
\end{eqnarray}
Combining these relations with (\ref{step2}) implies,
\begin{eqnarray}
\lefteqn{ \Bigl[ \square \!-\! (D \!-\! 1) H^2 \Bigr] D_{\rho} D^{[\mu}
\Bigl[ D^{\rho ]} {D'}^{\nu} y \!\times\! \frac{i}2 \gamma_s \Bigr] =
\Bigl( {\rm delta\ functions}\Bigr) } \nonumber \\
& & \hspace{4.5cm} + \frac{i e^2 H^2}{4 \pi^2 s} \, D^{\mu} {D'}^{\nu}
\Bigl[ A( y_{\scriptscriptstyle ++}) \!-\! A(y_{\scriptscriptstyle +-})
\Bigr] \; . \qquad
\end{eqnarray}
The fact that the left hand side is transverse fixes the delta functions
on the right hand side,
\begin{eqnarray}
\lefteqn{ \Bigl[ \square \!-\! (D \!-\! 1) H^2 \Bigr] D_{\rho} D^{[\mu}
\Bigl[ D^{\rho ]} {D'}^{\nu} y \!\times\! \frac{i}2 \gamma_s \Bigr] }
\nonumber \\
& & \hspace{2cm} = -\frac{e^2 H^2}{4 \pi^2 s} \Biggl\{ g^{\mu\nu}
\frac{ \delta^4(x \!-\! x')}{\sqrt{-g}} \!-\! D^{\mu} {D'}^{\nu}
\Bigl[ i A(y_{\scriptscriptstyle ++}) \!-\! i A(y_{\scriptscriptstyle +-})
\Bigr] \Biggr\} \; . \qquad
\end{eqnarray}
We can now recognize the right hand side as the transverse
projection operator \cite{TW2}, which means that relation
(\ref{Proca}) is indeed correct.

\section{Discussion}\label{discuss}

Even on flat space background, quantum corrections to the vacuum
polarization exhibit fascinating and subtle phenomena such as the
running of coupling constants. Corrections from quantum gravity
\cite{LW} even manifest the long-conjectured smearing of the
light-cone \cite{smear}. These sorts of quantum effects derive from
the response of classical electromagnetism to the ensemble of
virtual particles called forth by the Uncertainty Principle. One
might expect even stronger quantum effects on de Sitter background
because the inflationary expansion of spacetime rips gravitons and
light charged scalars out of the vacuum. The correction from charged
scalars on de Sitter is so large that the photon develops a mass
\cite{PTW,PP}, and comparably strong modifications are expected to
the electrodynamic force. The effect from dynamical gravitons is
expected to make the photon field strength grow with time, in
analogy to what happens for fermions \cite{MW}.

The focus of this paper has been on how to represent the tensor
structure of the vacuum polarization on de Sitter background. The
original studies of one loop corrections from scalar QED employed
a noncovariant form \cite{PTW,PP},
\begin{equation}
i \Bigl[ \mbox{}^{\mu} \Pi^{\nu}\Bigr](x;x') = \partial_{\rho}
\partial'_{\sigma} \Bigl\{ \eta^{\mu [\nu} \eta^{\sigma ] \rho}
\!\times\! F(x;x') \!+\! \overline{\eta}^{\mu [\nu}
\overline{\eta}^{\sigma ] \rho} \!\times\! G(x;x') \Bigr\} \;,
\label{oldrep}
\end{equation}
where $\eta^{\mu\nu}$ is the spacelike Minkowski metric and
$\overline{\eta}^{\mu\nu} \equiv \eta^{\mu\nu} + \delta^{\mu}_0
\delta^{\nu}_0$ is its purely spatial part. In section~\ref{MMC}
we developed a covariant extension (\ref{ansatz}-\ref{generic_tf})
using covariant derivatives, rather than ordinary ones, and
antisymmetrized products of the natural basis tensors
(\ref{basis}). When the physics breaks de Sitter invariance,
but preserves homogeneity and isotropy, it is possible to
represent the vacuum polarization with two structure functions,
just like the old representation (\ref{oldrep}).

For the special case where the vacuum polarization is physically de
Sitter invariant, one can employ a de Sitter invariant
representation (\ref{gammarep}) based on just one structure
function. The vacuum polarization from a massive scalar is de Sitter
invariant \cite{PP} and we worked out the fully renormalized structure
function (\ref{gammaren}). However, we saw in section~\ref{dsinv}
that there does not seem to be any advantage to using this
representation over the original one (\ref{oldrep}). With the old
formalism one can rather simply show that the leading contribution,
for small scalar mass, is a local Proca term \cite{PP}. In the de
Sitter invariant formalism this same result appears in the
cumbersome form of Green's Second Identity,
\begin{eqnarray}
\lefteqn{ - \frac{e^2 H^2}{4 \pi^2 s} \, g^{\mu\nu} A_{\nu}(x) =
\int \!\! d^4x' \Bigl[ \mbox{}^{\mu} G_s^{\nu}\Bigr](x;x') \Bigl[
\square' \!-\! (D \!-\! 1) H^2\Bigr] A_{\nu}(x') } \nonumber \\
& & \hspace{0cm} - \int_{\eta_i} \!\!\! d^3x' \Biggl\{ {D'}^0 \Bigl[
\mbox{}^{\mu} G_s^{\nu}\Bigr](x;x') \!\times\! A_{\nu}(x') - \Bigl[
\mbox{}^{\mu} G_s^{\nu} \Bigr](x;x') \!\times\! {D'}^0 A_{\nu}(x')
\Biggr\} \; , \qquad \label{dumb}
\end{eqnarray}
where $[\mbox{}^{\mu} G_s^{\nu}](x;x')$ is,
\begin{equation}
\Bigl[ \mbox{}^{\mu} G_s^{\nu} \Bigr](x;x') = \frac{i e^2}{16 \pi^4 s} \,
D_{\rho} D^{[\mu} \Biggl[ D^{\rho]} {D'}^{\nu}y \!\times\! \Bigl\{
\frac{ \ln(\frac{y_{\scriptscriptstyle ++}}4)}{4 \!-\!
y_{\scriptscriptstyle ++}} - \frac{ \ln(\frac{y_{\scriptscriptstyle ++}}4)}{
4 \!-\! y_{\scriptscriptstyle ++}} \Bigr\} \Biggr] \; . \label{dumber}
\end{equation}
It is difficult to discern any advantage the right hand side of
(\ref{dumb}) might possess over the left hand side, no matter how
passionately devoted one is to de Sitter invariance.

The presence of initial time surface integrals in expression (\ref{dumb})
is particularly disturbing. Long experience with the noncovariant
formalism has conditioned us to ignore such surface integrals because
they are likely to redshift rapidly, and because they can sometimes
be absorbed into perturbative corrections to the initial state
\cite{PTW,PP,PW,PTsW,KW,OW,FW,PW2,BOW,DW,MW,KW2,SPW1,SPW2,LW,KOW}.
As we proved in section~\ref{dsinv}, the surface integrals of (\ref{dumb})
cannot be dropped. Employing the de Sitter invariant formalism
would therefore require a painstaking study of the distinction
between surface terms which can and cannot be absorbed into
state corrections.

A major motivation for this study was to determine the best way
of representing the quantum gravitational contribution to the
vacuum polarization. Based on our results, there seems little
point to employing the covariant representation
(\ref{ansatz}-\ref{generic_tf}). The graviton propagator breaks
de Sitter invariance
\cite{TW1,RPW,MTW4,MTW2,MTW3,MTW1,KMW,MTW}, so the covariant structure
functions would not be de Sitter invariant in any case. But the
larger problem is that they would not be {\it simple}, nor would
they provide a simple picture of the physics. We therefore
conclude the old representation (\ref{oldrep}) is best. Should
this judgement prove incorrect, nothing essential has been lost
because the formalism of section~\ref{MMC}, plus some identities
for using the basis tensors (\ref{basis}) to express
$\eta^{\mu\nu}$ and $\overline{\eta}^{\mu\nu}$ \cite{KMW,PMW},
would allow us to convert the old representation (\ref{oldrep})
into the covariant form (\ref{ansatz}-\ref{generic_tf}).

Our work has some larger implications. One of these concerns
representing the vacuum polarization on a general background. A
plausible general representation for the vacuum polarization is
the same as our ansatz (\ref{ansatz}), but with the tensor
$[\mbox{}^{\mu\rho} T^{\nu\sigma}](x;x')$ expressed using
derivatives of the geodetic length-squared $\ell^2(x;x')$,
\begin{eqnarray}
\lefteqn{\Bigl[\mbox{}^{\mu\rho} T^{\nu\sigma}\Bigr](x;x') =
D^{\mu} {D'}^{[\nu} \ell^2 \, {D'}^{\sigma]} D^{\rho} \ell^2
\!\times\! f_1(x;x') } \nonumber \\
& & \hspace{5cm} + D^{[\mu} \ell^2 \, D^{\rho]} {D'}^{[\nu} \ell^2 \,
{D'}^{\sigma]} \ell^2 \!\times\! f_2(x;x') \; . \qquad
\end{eqnarray}
We expect that two structure functions should be needed --- as
they are for the massless, minimally coupled scalar on de Sitter
--- because classical electromagnetism shows birefringence for a
general metric \cite{Beilok}.

Another implication concerns the closely related problem of
representing the tensor structure of the graviton self-energy on de
Sitter background, $-i[\mbox{}^{\mu\nu} \Sigma^{\rho\sigma}](x;x')$.
The one loop contribution from massless, minimally coupled scalars
has been computed for noncoincident points \cite{PNRV}, and does not
seem especially complicated. However, when the required six
derivatives are extracted from this in a de Sitter invariant
fashion, so that the result can be renormalized and then employed to
quantum-correct the linearized Einstein equations, the two structure
functions require a full page to display \cite{SPW2}! The majority
of this complication derives from insisting on a de Sitter invariant
representation. That is not even possible for the contribution from
gravitons \cite{TW4}. We therefore conclude that a simple,
noncovariant representation for the graviton self-energy is
preferable.

\vskip 1cm

\centerline{\bf Acknowledgements}

This work was partially supported by FQXi Mini-Grant MGA-1222,
by FOM grant FP 104, by NSF grant PHY-1205591, and by the Institute
for Fundamental Theory at the University of Florida.

\end{document}